\documentclass[%
 aip,
 pop,
 amsmath,amssymb,
 reprint,%
]{revtex4-2}

\usepackage{graphicx}
\usepackage{dcolumn}
\usepackage{bm}

\usepackage[utf8]{inputenc}
\usepackage[T1]{fontenc}
\usepackage{mathptmx}
\usepackage{soul}
\usepackage{etoolbox}
\usepackage[colorlinks=true,linkcolor=blue,citecolor=blue,urlcolor=blue]{hyperref}
\usepackage[none]{hyphenat}
\usepackage{multirow}
\usepackage[normalem]{ulem}
\begin{document}

\preprint{AIP/123-QED}

\title{Revisiting Electron Heating in Capacitively Coupled Plasma (CCP) Discharges: A Nonlinear Dynamics Perspective}
\author{Rishabh Singh}
\email{rishabh.singh@ipr.res.in}
\affiliation{Institute for Plasma Research, Bhat, Gandhinagar, Gujarat 382428, India}
\affiliation{Homi Bhabha National Institute, Anushaktinagar, Mumbai 400094, India}
\author{Sarveshwar Sharma}
\email{sarvsarvesh@gmail.com}
\affiliation{Institute for Plasma Research, Bhat, Gandhinagar, Gujarat 382428, India}
\affiliation{Homi Bhabha National Institute, Anushaktinagar, Mumbai 400094, India} 

\author{Bhooshan Paradkar}
\affiliation{School of Physical Sciences, UM-DAE Centre for Excellence in Basic Sciences, University of Mumbai, Mumbai 400098, India} 

\author{Sudip Sengupta}
\affiliation{Institute for Plasma Research, Bhat, Gandhinagar, Gujarat 382428, India}
\affiliation{Homi Bhabha National Institute, Anushaktinagar, Mumbai 400094, India}

\date{\today}

\begin{abstract}
Despite decades of research, the electron heating mechanisms in capacitively coupled plasma (CCP) discharges over a wide range of operating conditions is not fully understood. Although stochastic heating is generally regarded as the dominant collisionless heating mechanism at low pressures, the inherently nonlinear electron dynamics responsible for this process have not been fully quantified. These nonlinear interactions drive stochastic heating, a mechanism considered crucial for energy transfer in CCPs, yet its quantitative impact on plasma parameters remains insufficiently explored. In this work, we investigate electron dynamics in steady-state CCP discharges and demonstrate that electron motion in the plasma bulk exhibits intrinsically chaotic behavior. The onset of chaos is identified using Poincaré sections and quantified through Lyapunov exponent analysis. To further quantify this behavior, we map the spatial distribution of the Lyapunov exponent—normalized by the electron–neutral collision frequency—across the plasma bulk for different pressures and RF voltages. The normalized Lyapunov exponent increases systematically with decreasing pressure and increasing RF voltage, indicating enhanced stochasticity and a stronger sensitivity of electron trajectories to initial conditions. These results establish the Lyapunov exponent as a quantitative measure of effective stochastic scattering in collisionless CCPs and provide a direct comparison with the classical stochastic collision frequency proposed by Popov and Godyak [Journal of Applied Physics 57, 53–58 (1985)]. The present analysis offers a unified nonlinear dynamical framework for understanding stochastic electron heating in low-pressure RF plasmas.
 
\end{abstract}

\maketitle
\section{introduction}
\par Capacitvely coupled plasmas (CCP) are among the most technologically important plasma sources in semiconductor manufacturing, where they are widely used for plasma etching and thin-film deposition in large-scale integrated circuit fabrication \cite{Lieberman2005,Chabert2011}. Low-pressure plasma discharges can be sustained by radio-frequency (RF) currents and voltages applied to electrodes immersed in the plasma. Under these conditions, a high-voltage capacitive sheath forms between the electrode surface and the quasi-neutral bulk plasma, serving as the primary region of energy transfer between the oscillating electric field and charged particles.
\par Historically, CCP systems were predominantly operated at the standard industrial frequency of 13.56 MHz. More recently, operation at very high frequencies (VHF, 30–300 MHz) has attracted considerable attention because VHF discharges produce substantially higher plasma densities for a given discharge power. This increase arises primarily from enhanced current at higher frequencies, which enables efficient plasma generation while maintaining lower DC self-bias voltages. Such characteristics are particularly advantageous for large-area wafer processing and advanced device fabrication requiring reduced ion-induced surface damage\cite{rauf2010power, wilczek2015effect,sharma2018plasma,ss19}. VHF-driven CCP discharges also exhibit distinct physical phenomena not commonly observed at lower frequencies. These include the formation of transient electric field structures in the plasma bulk, nonlinear harmonic generation through plasma–sheath coupling, and the emergence of energetic electron beams during rapid sheath collapse\cite{upadhyay2013effect, miller2006spatial, Sharma2022,Sharma2016, wilczek2018disparity}. Despite these unique features, a comprehensive and quantitative understanding of electron heating mechanisms in VHF CCPs remains incomplete, particularly at the level of individual particle dynamics.
\par In low neutral gas pressure VHF regime, classical ohmic (collisional) heating—arising from electron–neutral momentum transfer—becomes inefficient for sustaining the discharge, and alternative mechanisms dominate the electron energy gain process. In this regime, electrons gain energy primarily through interactions with the oscillating plasma–sheath boundary, a process known as stochastic or collisionless heating. This mechanism has been extensively investigated through both theoretical and experimental studies over several decades \cite{Lieberman1988,Godyak1972,Kaganovich2002,Akhiezer1976,Surendra1991,Turner1995,Gozadinos2001b,Godyak1990,Godyak1976,Godyak1979,Popov1985}. 
\par The theoretical foundation of collisionless heating in CCPs was established by Godyak \cite{Godyak1976,GodyakKhanna1976}, who introduced one of the earliest explicit applications of Fermi acceleration to electron heating in RF discharges. In the so-called `hard wall model', the oscillating sheath is represented as a perfectly reflecting moving boundary, and electrons interacting with this boundary undergo elastic reflections that result in net energy gain. Using this framework, analytical estimates of power deposition were obtained for weakly modulated sheath potentials, demonstrating that Fermi acceleration can serve as a major mechanism for sustaining low-pressure discharges. These early models were further refined by several authors, including Akhiezer and Bakai \cite{Akhiezer1976}, who employed simplified Fermi-type models to derive analytical expressions for the electron heating rate. Experimental studies later confirmed the dominance of collisionless heating over electron–neutral collisional heating in sufficiently low-pressure regimes \cite{Godyak1976,GodyakKhanna1976,Popov1985}.
\par A more complete and self-consistent description of RF sheath dynamics was later developed by Lieberman \cite{Lieberman1988}, who derived analytical solutions for the coupled sheath–plasma system without relying on the simplifying assumptions inherent to the hard wall model. In Lieberman's formulation, electrons possess distinct densities at different sheath boundaries and carry finite drift velocities, yielding a more physically realistic description of the sheath structure. However, subsequent studies indicated certain limitations of this approach. It was later pointed out by Gozadinos \textit{et al.}\cite{Gozadinos2001a,Gozadinos2001b}, using a fluid approach, that Lieberman's model fails to satisfy current conservation at the electron sheath edge — a condition under which the instantaneous electron heating vanishes identically in the hard wall approximation. To incorporate kinetic effects more accurately, Kaganovich\cite{Kaganovich2002} employed a two-step ion density model that incorporates bulk plasma oscillations, which had been entirely neglected in the earlier hard wall formulations, thereby offering a more complete description of RF sheath dynamics at low pressures.
\par Beyond stochastic sheath interactions, additional electron heating channels have been identified. Surendra and Graves,\cite{Surendra1991} and later Turner,\cite{Turner1995} demonstrated the existence of pressure heating arising from spatial gradients in electron density and temperature near the moving sheath boundary. During sheath expansion, electrons at the plasma-sheath boundary are compressed, gaining thermal energy; during sheath collapse, rarefaction occurs. These compression and rarefaction cycles produce thermal disturbances that constitute a net energy input to the electron population, distinct from and complementary to the Fermi stochastic heating mechanism. Particle-in-cell (PIC) simulations performed by Kawamura \textit{et al.}\cite{Kawamura2006} and Sharma \textit{et al.}\cite{Sharma2013,SharmaThesis2013} enabled systematic comparison between competing heating mechanisms, significantly advancing the quantitative understanding of power deposition in CCP discharges. 
\par Although significant progress has been made in describing macroscopic power deposition mechanisms, the microscopic phase-space dynamics governing individual electron motion and its connection with heating processes is relatively less explored. A detailed investigation of the phase-space dynamics is crucial not only for a fundamental understanding of the underlying stochastic processes but also for their quantitative characterization. The stochasticity in low pressure CCP discharges comes from the fact that the electrons undergo repeated interactions with oscillating sheath boundaries, leading to progressive phase randomization relative to the RF field. This loss of phase coherence enables nonzero time-averaged energy absorption, whereas perfectly phase-coherent electron motion would result in symmetric energy exchange and zero net heating. As a consequence, electron trajectories in phase space do not remain periodic or closed but instead evolve into irregular patterns that depend sensitively on initial conditions.
\par This irregular, sensitive dependence of electron trajectories on initial conditions is precisely the hallmark of deterministic chaos in Hamiltonian dynamical systems.\cite{bak} To rigorously characterize this behavior, it is natural to employ the tools of nonlinear dynamics. The widespread adoption of PIC simulations for kinetic modeling of CCP discharges makes such investigations readily feasible. However, despite the rich phase-space information available in PIC data, to best of our knowledge,  no studies have leveraged it for the quantitative characterization of the underlying stochastic processes. The present work is an attempt in this direction. 
\par In particular, we utilize two standard and widely adopted nonlinear dynamical diagnostics: the Poincaré section\cite{bak} and the Lyapunov exponent.\cite{bak,strogatz} The Poincaré section provides a stroboscopic portrait of the electron phase space by recording the particle's position and velocity at regular intervals (once per RF period), revealing whether trajectories lie on invariant tori (regular motion), fill resonant island chains (near-integrable motion), or scatter diffusely across phase space (chaotic motion). The Lyapunov exponent quantifies the rate of exponential separation of initially neighboring trajectories in phase space: a positive Lyapunov exponent $\lambda_{L}>0$ is the definitive signature of chaos, confirming that the system is sensitive to initial conditions and that nearby trajectories diverge exponentially in time as\cite{bak} $|\delta x(t)|\approx |\delta x(0)|e^{\lambda_L t}$. Together, these two diagnostics provide a comprehensive, quantitative characterization of the stochastic nature of electron motion in CCP discharges. Specifically, using these diagnostics over a phase-space data, obtained from PIC simulations carried over wide range of applied voltages and neutral gas pressures, we address the issue of transition between collisional and collisionless electron heating. 
\par The remainder of the paper is organized as follows. Section II describes the simulation methodology, including the numerical technique, choice of parameters, and implementation of boundary conditions. Section III presents the test-particle simulation framework, along with the construction of Poincaré sections, evaluation of Lyapunov exponents, and the methodology used to quantify electron heating. Finally, Section IV summarizes the principal findings of this study and provides concluding remarks.
\section{simulation technique }
\par In this work, we employ the one-dimensional-in-space, three-dimensional-in-velocity
(1D–3V) Electrostatic Direct Implicit Particle-In-Cell (EDIPIC) code\cite{sydorenko2006particle,campanell2012instability,campanell2013influence,carlsson2017validation,charoy20192d,sheehan2013kinetic,ss18,PhysRevResearch.4.013059,Sharma2022} to simulate the
plasma discharge. EDIPIC is a well-established and extensively benchmarked simulation
platform that has undergone rigorous verification against analytical models and validation
against experimental measurements,\cite{godyak1992measurement,Sharma2022} making it a reliable and well-suited tool for the
investigation of low-pressure capacitively coupled plasma discharges. Although the code is
designed to operate in both explicit and implicit time-integration modes, all simulations
presented in this work are performed exclusively in the explicit mode.
\par The physical model describes a plasma bounded between two infinite, parallel-plate
electrodes and is formulated within the well-established particle-in-cell/Monte Carlo collision
(PIC–MCC) framework \cite{birdsall2018plasma, hockney2021computer}. The collision physics incorporated in the model is comprehensive for the pressure regime
under consideration. For electron–neutral collisions, three processes are included: elastic
(momentum transfer) scattering, electron excitation, and ionization. For ion–neutral
collisions, elastic scattering and resonant charge-exchange are taken into account. The plasma
chemistry model includes reactions associated with the first excited state of argon (11.6 eV).
However, the metastable species itself is not explicitly tracked as a separate particle
population in the simulation, as their contribution to the overall discharge dynamics is
negligible within the operating parameter range considered here and would not materially
affect the results. All collision cross-section data employed in the calculations are sourced
from well-established, experimentally validated databases,\cite{rauf1997argon,lauro2004analysis} ensuring the physical
accuracy of the collision operator. Secondary electron emission from the electrode surfaces has been neglected in the present
model. This simplification is justified by the low-pressure operating conditions considered
here, under which the ion bombardment energies and resulting secondary electron emission
yields are sufficiently small to neglect their influence on the global discharge characteristics \cite{wen2023importance, horvath2018effect}.
\par The external RF circuit is not explicitly included in the present simulations. Instead, a fixed
sinusoidal voltage waveform is imposed directly on the powered electrode, while the
grounded electrode is held at zero potential throughout the simulation. This approach, which
is standard practice in PIC–MCC simulations of symmetric CCP discharges, allows the self-
consistent current and charge density profiles to develop naturally in response to the applied
potential, without the additional complexity introduced by circuit elements such as blocking
capacitors or matching networks. Under these conditions, a DC self-bias develops self-
consistently as a consequence of the difference in electron and ion mobilities.
\par
\begin{figure}[!htbp]
 \centering
 \includegraphics[width=1.0\linewidth]{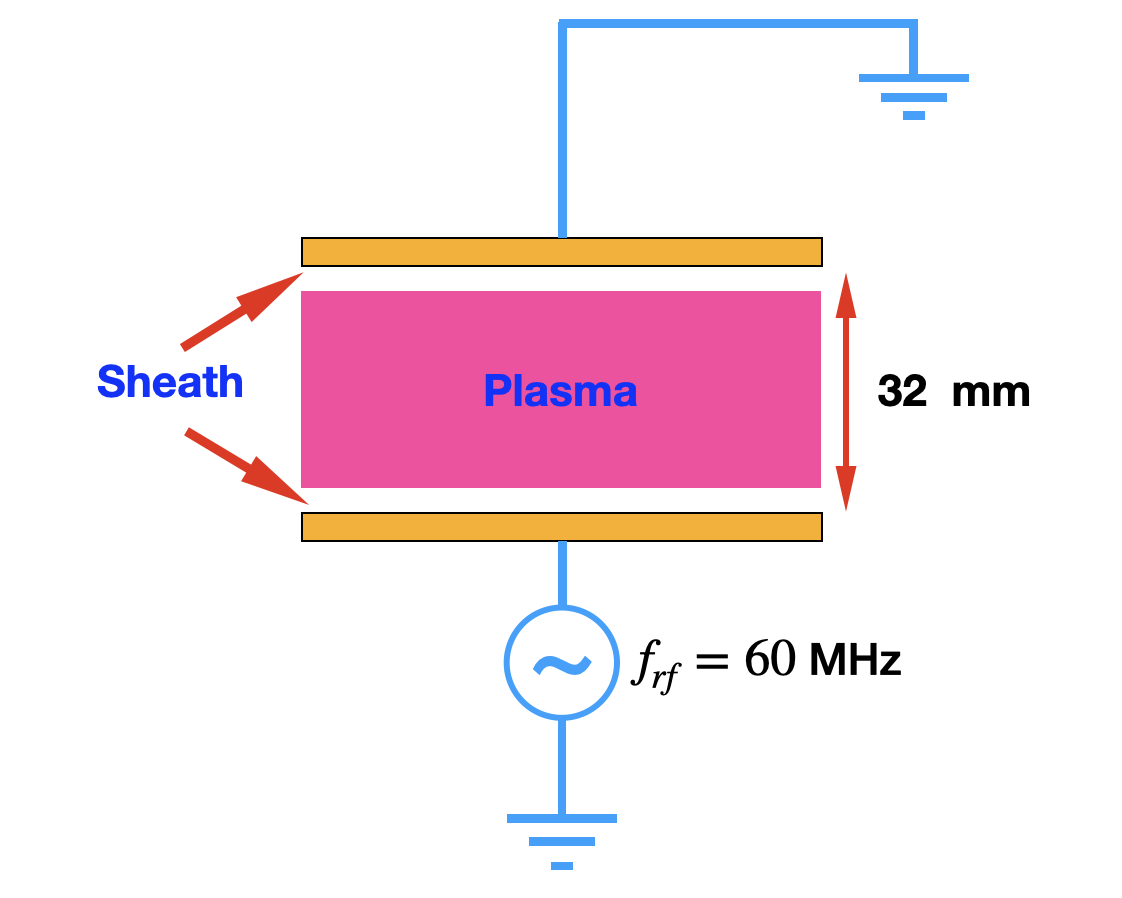}
 \caption{A schematic diagram of the CCP discharge.}
 \label{smtk}
\end{figure}

 A schematic diagram of the CCP discharge is shown in Figure\ref{smtk}.
The sinusoidal voltage waveform applied to the powered electrode takes the standard form:
 \begin{equation}
     V(t) = V_0 \sin (2\pi f_{rf}t + \phi)
 \end{equation}
where $V_0$ is the  applied voltage amplitude, $f_{rf}$ is the driving frequency and $\phi$ is the phase,
while the grounded electrode is maintained at 0 V throughout the simulation. The resulting asymmetric potential boundary conditions drive the plasma discharge and give rise to the self-consistent sheath dynamics, electron heating and higher harmonic generation that are the primary focus of this study. In the present simulations, the initial phase of the applied sinusoidal voltage waveform is set to $\phi=0$. This choice of initial phase is conventional and does not affect the steady-state discharge characteristics. The driving frequency is held constant at 60 MHz throughout all simulations, placing the discharge firmly in the VHF regime. The system length L, defined as the physical separation between the powered electrode (PE)
and the grounded electrode (GE), is fixed at 32 mm. This inter-electrode gap is sufficiently large to accommodate well-developed sheath regions at both electrodes and an extended quasi-neutral bulk plasma in between, while remaining small enough that the one-dimensional approximation is fully justified given the electrode dimensions. 
\par The neutral argon gas pressure is varied systematically over the range of 5–100 mTorr (i.e. 5, 30, 50, 70 and 100 mTorr), spanning the transition from highly collisionless to moderately collisional discharge conditions. The applied RF voltage amplitude $V_0$ is independently varied between 5 V and 100 V (i.e. 5, 30, 50, 70 and 100 V), enabling a systematic investigation of the role of the driving voltage amplitude on discharge characteristics. The initial ion temperature is set to $T_i$ = 0.026 eV, corresponding to room temperature ($\approx$300 K), consistent with the assumption that ions are in thermal equilibrium with the neutral gas at the start of the
simulation. The initial electron temperature is set to $T_e$ = 1.5 eV, a value typical of the early phase of discharge ignition in low-pressure RF plasmas, from which the electron population self-consistently evolves to its steady-state energy distribution under the influence of the applied field and collisional processes.
\par The spatial grid cell size is chosen to be one-eighth of the electron Debye length i.e. $\Delta x = \lambda_{De}/8$, to ensure that the fine-scale features in plasma density and potential are accurately resolved. Here, plasma density of $n_0$ = $5  \times 10^{15}\,m^{-3} $ is used for calculation of $\lambda_{De}$.  The time step is chosen as per the CFL condition $\Delta t= \Delta x/{v_{max}}$, where maximum velocity is taken as four times the thermal velocity i.e. $v_{max}= 4\times v_{th,e}$. This conservative estimate ensures that even the most energetic electrons in the high-energy tail of the distribution function are correctly resolved temporally and do not traverse more than one grid cell per time step — a necessary condition for the numerical stability and physical accuracy of the explicit PIC algorithm. This choice of timestep also ensures that the fastest collective oscillations in the plasma, namely the electron plasma oscillations at frequency $\omega_{pe}$, are resolved with high temporal fidelity and that the equations of motion are integrated in a numerically stable and
physically accurate manner throughout the entire simulation duration. Satisfying this criterion also prevents the well-known numerical artifact of artificial electron heating, which arises in explicit PIC simulations when $\omega_{pe}\Delta t$ approaches or exceeds the stability boundary ($\omega_{pe}\Delta t < 0.2$), leading to spurious energy injection into the particle distribution and non-physical modifications of the electron energy distribution function.
\par At the start of each simulation, the computational domain is uniformly populated with 256
super-particles per grid cell for each species — electrons and singly ionized argon ions ($Ar^+$)
— resulting in a total initial particle count of roughly less than half a million computational
particles per species across the 1406-cell grid. These particles are subjected to perfectly absorbing boundary conditions at both electrode surfaces. Thus, when a charged particle — electron or ion — reaches either the powered or grounded electrode during the course of its trajectory, it is immediately removed from the
simulation domain and its charge is deposited on the corresponding electrode, contributing to the instantaneous electrode current. The self-consistent charge balance between the two electrodes — including the development of a DC self-bias — emerges naturally from the asymmetric flux of electrons and ions to the electrode surfaces under these boundary conditions.  
\par To ensure that all diagnostic quantities are extracted from a physically meaningful, fully
developed steady-state discharge, each simulation is carried out for a duration exceeding
5000 RF cycles. At the operating frequency of 60 MHz, one RF cycle corresponds to a period
of $T_{RF}$ = $1/f_{rf} \approx16.67$ ns, so the total simulation time spans more than 83 $\mu s$. This extended
simulation duration is necessary because the discharge must traverse a complex transient
phase — involving plasma buildup from the initial seed density, sheath formation, and the
self-consistent adjustment of the plasma potential, electron temperature, and ion density
profiles — before reaching a statistically stationary state in which all cycle-averaged
quantities are stable. The attainment of steady state is confirmed by monitoring the time
evolution of cycle-averaged plasma density, mean electron energy, and electrode current, and
verifying that these quantities exhibit no systematic drift over the final 500 RF cycles
from which all diagnostic averages are computed. This rigorous convergence criterion
ensures that the simulation results are representative of the true steady-state discharge
behavior and are free from transient initialization artifacts.
\section{Results} 
\par The dynamical response of test electrons critically depends upon these steady-state discharge characteristics such as plasma density, electric field/plasma potential, bulk plasma length etc. Therefore, before analyzing the phase-space portrait of test electrons, we first discuss these steady-state discharge characteristics. These characteristics are obtained by averaging over 100 RF cycles after the steady-state is reached (typically after 5000 RF cycles).  
\par
\begin{figure*}
 \centering
 \includegraphics[width=\textwidth]{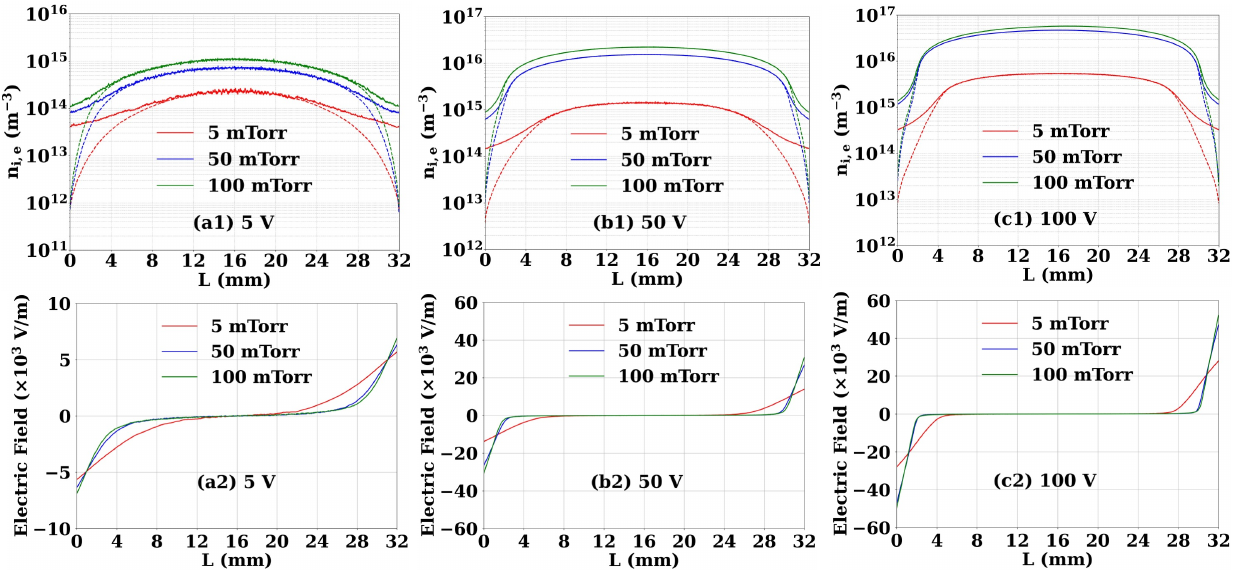}
  \caption{In this figure (a1), (b1) and (c1) are  time averaged ion density ($n_i$, solid line) and electron density ($n_e$, dashed line). (a2), (b2) and (c2) are time averaged electric field at 5 V, 50 V and 100 V respectively for 5, 50 and 100 mTorr.}
 \label{avd}
\end{figure*}

 Figure \ref{avd} shows the spatial profiles of steady-state plasma density((a1)–(c1)) and electric field ((a2)–(c2)) for the applied RF voltages of 5 V, 50 V, and 100 V. These studies are performed at gas pressures of 5 mTorr (red), 50 mTorr (blue), and 100 mTorr (green). The ion and electron densities for these conditions are plotted with solid and dashed lines, respectively. The steady-state characteristic such as peak plasma density ($n_0$), size of the bulk plasma ($L_{bulk}$) and maximum plasma potential ($V_p$) are listed in the table ~\ref{tab}.  From these results, we observe that at constant applied voltage, increasing the neutral gas pressure from 5 mTorr to 100 mTorr leads to a systematic increase in both the ion and electron densities, accompanied by a reduction in sheath width. Note that the sheath width can be estimated from the region where electric field exhibits spatial gradients (see Fig.\ref{avd}(a2)–(c2)). Conversely, at constant gas pressure, the plasma density exhibits significant enhancement with increasing applied voltage. For example, as $V_0$ is increased from 5 V to 100 V, the plasma density increases by factors of approximately 27, 67, and 52 at 5 mTorr, 50 mTorr, and 100 mTorr, respectively. Similar enhancement factors for increase in pressure from 5 mTorr to 100 mTorr at 5 V, 50 V and 100 V are approximately 6, 16 and 11, respectively.  
\begin{table}[h]
    \centering
    \begin{tabular}{|c|c|c|c|c|c|}
    \hline
      P (mTorr) & $V_{0}$ (V) & $n_0$ ( $\times 10^{15}$ m$^{-3}$)& $L_{bulk}$ (mm) & $V_{p}$ (V) & $V_{0}/V_{p}$ \\
      \hline
      \multirow{3}{*}{5} 
        & 5 & 0.2 & 12.5 & 26.1 & 0.2 \\
      \cline{2-6}
        & 50 & 1.4 & 17.8 & 40.9 & 1.2 \\
      \cline{2-6}
        & 100 & 5.4 & 21.8 & 63.2 & 1.6 \\
      \hline

      \multirow{3}{*}{50} 
        & 5 & 0.7   & 20.6 & 17.5 & 0.3 \\
      \cline{2-6}
        & 50 & 15.3 & 26.3 & 32.9 & 1.5 \\
      \cline{2-6}
        & 100 & 47.0 & 27.2 & 52.4 & 1.9 \\
      \hline

      \multirow{3}{*}{100} 
        & 5 & 1.1  & 23 & 16.7 & 0.3 \\
      \cline{2-6}
        & 50 & 22.2 & 27.1 & 32.9 & 1.5 \\
      \cline{2-6}
        & 100 & 57.6 & 27.4 &52.5 & 1.9 \\
      \hline
    \end{tabular}
    \caption{Steady-state discharge characteristics for various gas pressures and applied RF voltages ($V_0$). The peak plasma density, bulk plasma size and maximum plasma potential are represented by $n_0$, $L_{bulk}$ and $V_p$, respectively.}
    \label{tab}
\end{table}
\begin{figure*}[!htbp]
 \centering
 \includegraphics[width=13 cm, height=6.5 cm]{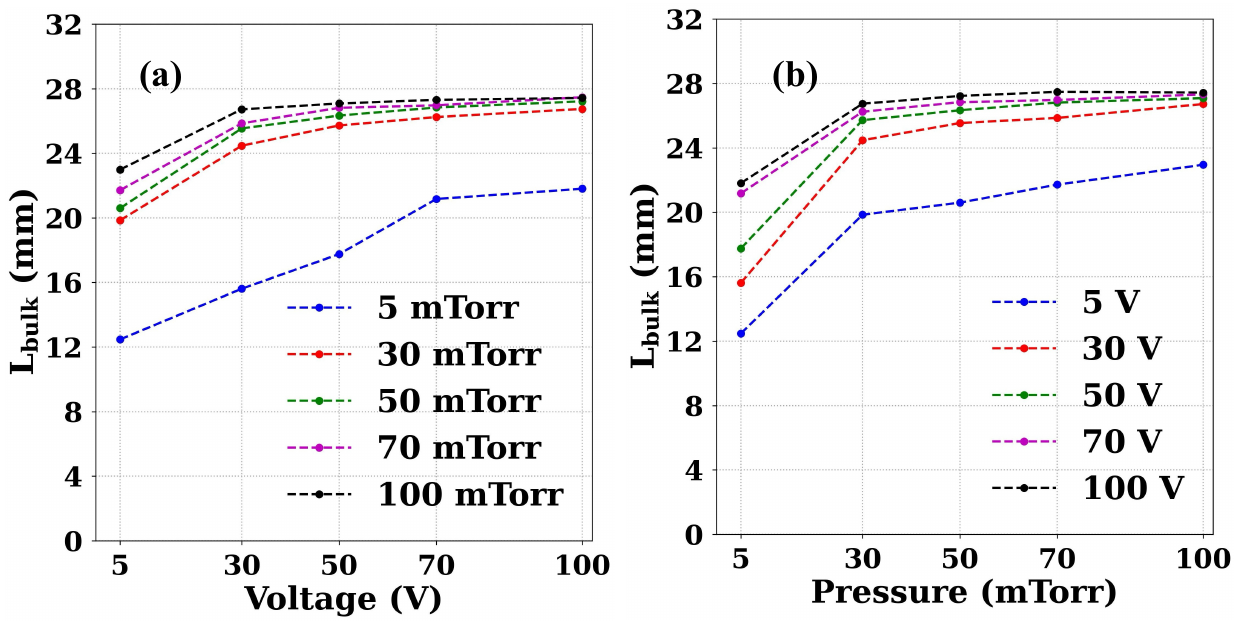}
  \caption{In this figure, (a) is variation of bulk length with voltage (5 to 100 V) at constant pressure (5 mTorr (blue), 30 mTorr (red), 50 mTorr (green), 70 mTorr (magenta) and 100 mTorr (black)), (b) is variation of bulk length with pressure (5 to 100 mTorr) at constant voltage (5 V (blue), 30 V (red), 50 V (green), 70 V (magenta), and 100 V (black))}
 \label{sw}
\end{figure*}
\par The size of bulk plasma, where $n_i \simeq n_e$, can be estimated from the electric field profile. The sheath width is obtained from  snapshots of the electron density as the maximum distance between the electrode and the electron sheath edge. The bulk length is then calculated as the total system length minus the sheath widths on both sides of the electrodes. The spatial gradient in the electric fields indicate violation of quasi-neutrality. Based on these consideration, the bulk plasma width ($L_{bulk}$) is estimated. Figure \ref{sw} and Table \ref{tab} show dependence of $L_{bulk}$ on the applied RF voltage and neutral gas pressures. Here, Figure \ref{sw}(a) and (b) show the trend in $L_{bulk}$ at constant pressure and voltage, respectively. Overall, these results indicate that the bulk plasma width is highly sensitive to both applied voltage and neutral gas pressure in the low-voltage and low-pressure regimes, whereas a tendency towards saturation emerges at higher operating conditions. This behavior reflects the competing effects of enhanced ionization, sheath contraction, and increased collisionality, which collectively determine the spatial extent of the quasi-neutral plasma bulk.
\par The physical origin of the bulk width expansion in both cases is rooted in enhanced ionization. At constant voltage, increasing the neutral gas pressure raises the neutral particle density, thereby increasing the electron–neutral collision frequency and the total rate of ionization events per unit volume. The resulting increase in plasma density compresses the sheath and expands the quasineutral bulk. At constant pressure, increasing the applied voltage elevates the mean electron energy and enhances the interaction rate between electrons and the oscillating sheath boundary. As electrons gain energy more rapidly through sheath-electric-field heating, the ionization rate of neutral species increases, driving a further expansion of the bulk plasma. Both mechanisms are therefore manifestations of intensified ionization, mediated either through collisional frequency (pressure-driven) or electron energy (voltage-driven).  
 \begin{figure*}
 \centering
 \includegraphics[width=\textwidth]{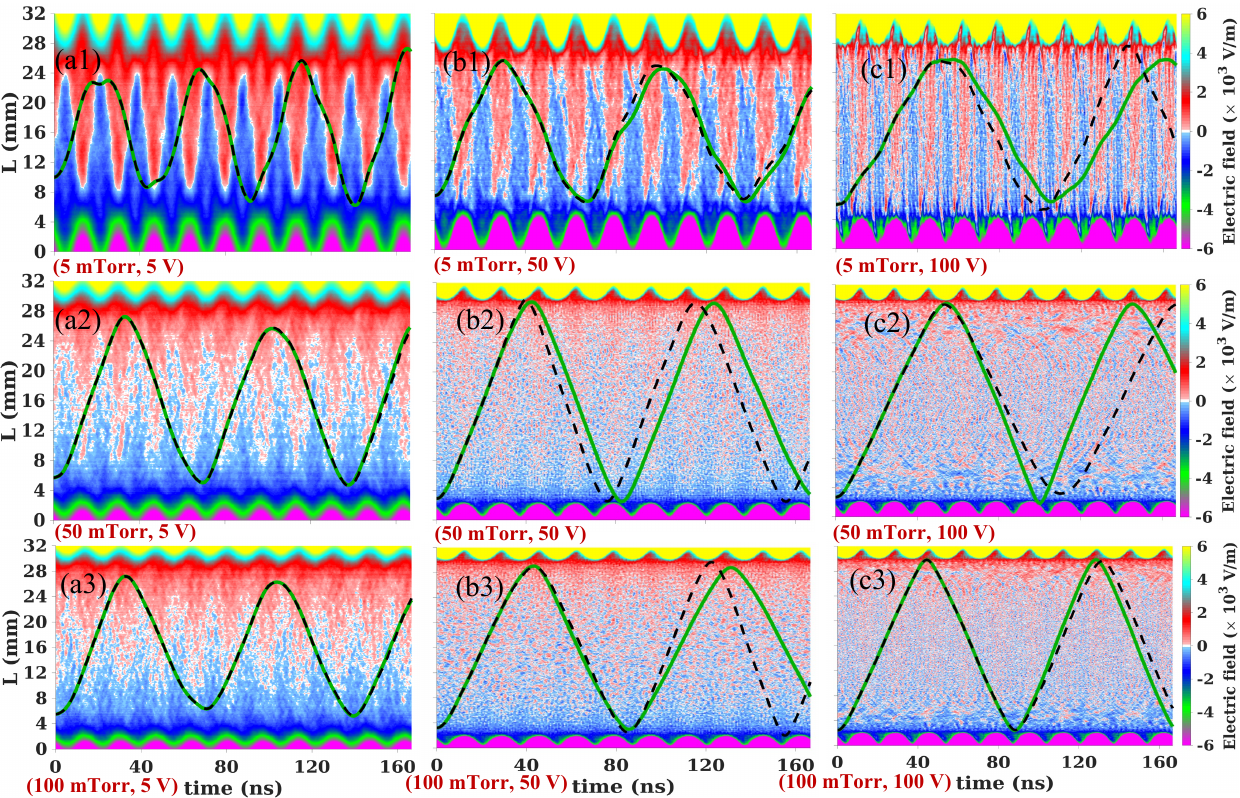}
  \caption{In this figure solid green ($x_1$) and dashed black ($x_2$) are trajectories of 0 eV test electron for 10 rf cycle placed near the sheath, both are initially separated by $\lambda_{De} /120$ . (a1) to (a3) are at 5 V, (b1) to (b3) are at 50 V and (c1) to (c3) are at 100V for 5, 50 and 100 mTorr.}
 \label{stp}
\end{figure*} 

\begin{figure*}
 \centering
 \includegraphics[width=\textwidth, height= 9 cm, keepaspectratio]{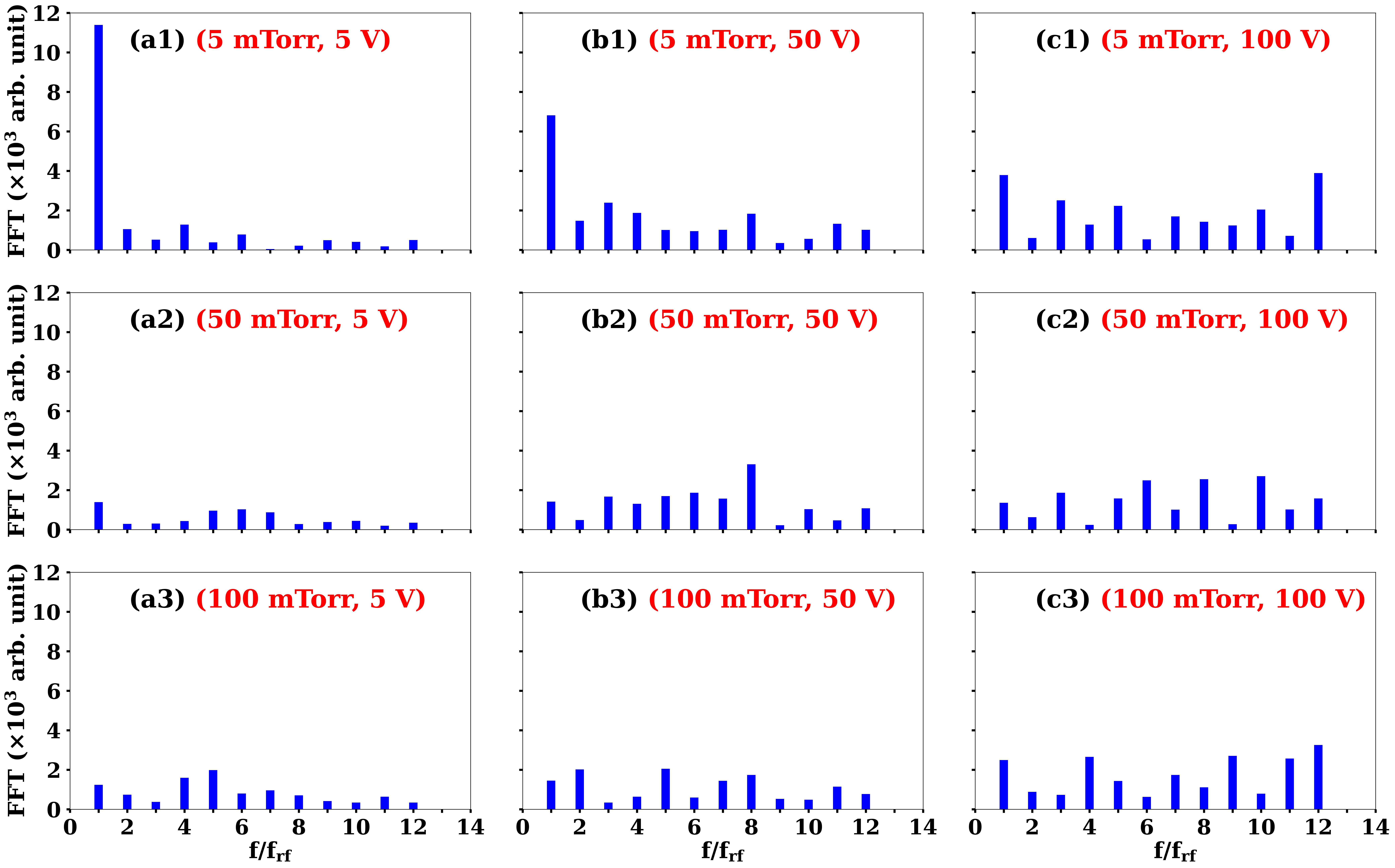}
  \caption{ FFT of Electric Field at the center of the discharge. (a1) to (a3) are at 5 V, (b1) to (b3) are at 50 V and (c1) to (c3) are at 100V for 5, 50 and 100 mTorr.}
 \label{fft}
\end{figure*}
\subsection*{Probing stochastic electric field with Test Particle Simulations}
\par Having established steady-state discharge characteristics, we now discuss the stochastic motion of test electrons under the steady-state conditions. For this, we investigate spatiotemporal profiles of electric field along with the trajectories of test electrons as shown in Figure \ref{stp}. Here we have plotted the electric field over 10 consecutive RF cycles following the attainment of steady state, for the complete set of voltage and pressure combinations explored in this study. To render the spatial structure of the bulk plasma features visible without saturation by the dominant sheath fields, the electric field magnitude has been clipped to the range -6 to +6 kV/m in all panels. Additionally, in Fig.~\ref{fft}, we have plotted the harmonics present in the electric field at the center of the discharge using the Fast Fourier Transform (FFT). The spatiotemporal maps reveal the presence of pronounced electric field transients extending across the bulk plasma, most clearly at low gas pressures. These transient structures originate at one sheath boundary and propagate across the full discharge length to reach the opposite sheath. \cite{Kaganovich2002,ig2006, ss13, sss19, ss19}    
\par Figures \ref{stp}(a1)–(c1), \ref{stp}(a2)–(c2), and \ref{stp}(a3)–(c3) correspond to neutral gas pressures of 5 mTorr, 50 mTorr, and 100 mTorr, respectively, with each row spanning applied voltages of 5 V, 50 V, and 100 V. In figure \ref{stp}(a), the bulk plasma exhibits distinct filamentary structures in the spatio-temporal electric fields at low pressure and low voltage (5 mTorr, 5V) operating conditions. These filaments are originated at the fundamental RF frequency, as can be seen from the dominance of fundamental harmonic in the Fig.~\ref{fft}(a1). With increasing voltage and pressure, these structures get diminished inside the bulk plasma. The breaking up of lumped filamentary structures with increasing applied voltage is due to enhanced non-linearity as can be seen from the presence of higher harmonics (up to 12$^{th}$ harmonic) in Fig.~\ref{fft}(b1,c1). Such type of filamentation with increasing applied RF voltage is reported in literature\cite{ss19}. On the contrary, enhanced electron-neutral collisions with increasing gas pressure can lead to suppression of formation of filaments, which are mainly driven by release of energetic electrons during the sheath expansion phase \cite{ig2006}. Therefore, with increasing pressure the breaking of filaments causes reduction in strength of the fundamental harmonic and generation of higher harmonics (see Fig.~\ref{fft}). Effect of gas pressure on higher harmonic generation in CCPs is reported in literature\cite{ss24}. It should be noted that at 100 mTorr pressure, filamentary structures are predominantly present only when the applied voltage is 5V. However, at higher neutral gas pressure conditions the electric field inside the bulk plasma gets considerably smoothened due to breaking of these filaments , as can be seen from Fig.~\ref{stp}. Thus, two competing processes, viz. enhancement in non-linearity with increasing applied voltage and breaking of transient filamentary structures with increasing pressure, set the instantaneous electric fields inside the bulk plasma. 
\par The stochastic nature of the electric field can be probed using the test electron trajectories, which are superimposed on the spatiotemporal plots of electric fields given in Fig.\ref{stp}. In each of the case, near left sheath edge, two test electrons, initially at rest and separated from each other by a distance $\lambda_{De}/120$, are tracked for 10 RF cycles shown in the figure. The stochastic electric field can then be characterized by the extent of separation between these two electrons as a function of time. For example, in 5 mTorr case (Fig.\ref{stp}(a1)-(c1)) we see that the trajectories of these two test electrons, shown by solid green and dashed black lines, become increasingly divergent with increasing applied voltage. On the other hand, trajectories are relatively less divergent for 100 mTorr case as can be seen Fig.\ref{stp}(a3)-(c3). These observations establish a clear qualitative correspondence between the electric field transients and the degree of stochasticity in it through the electron dynamics. In particular, discharge conditions that produce stronger, thinner, and more fragmented transient electric field structures also lead to more rapidly diverging electron trajectories. This indicates that the nonlinear electric field evolution in the bulk plasma plays a central role in governing the onset of irregular and potentially chaotic electron motion. 
\par In the following subsection, tools from nonlinear dynamics, namely phase-space portraits, Poincare sections and Lyapunov exponents, are employed to quantitatively characterize these oscillations and to systematically examine the degree of stochasticity in electron motion within the bulk plasma.  

\subsection{Phase Space Portrait}
\begin{figure*}
 \centering
 \includegraphics[width=\linewidth, height=11cm, keepaspectratio]{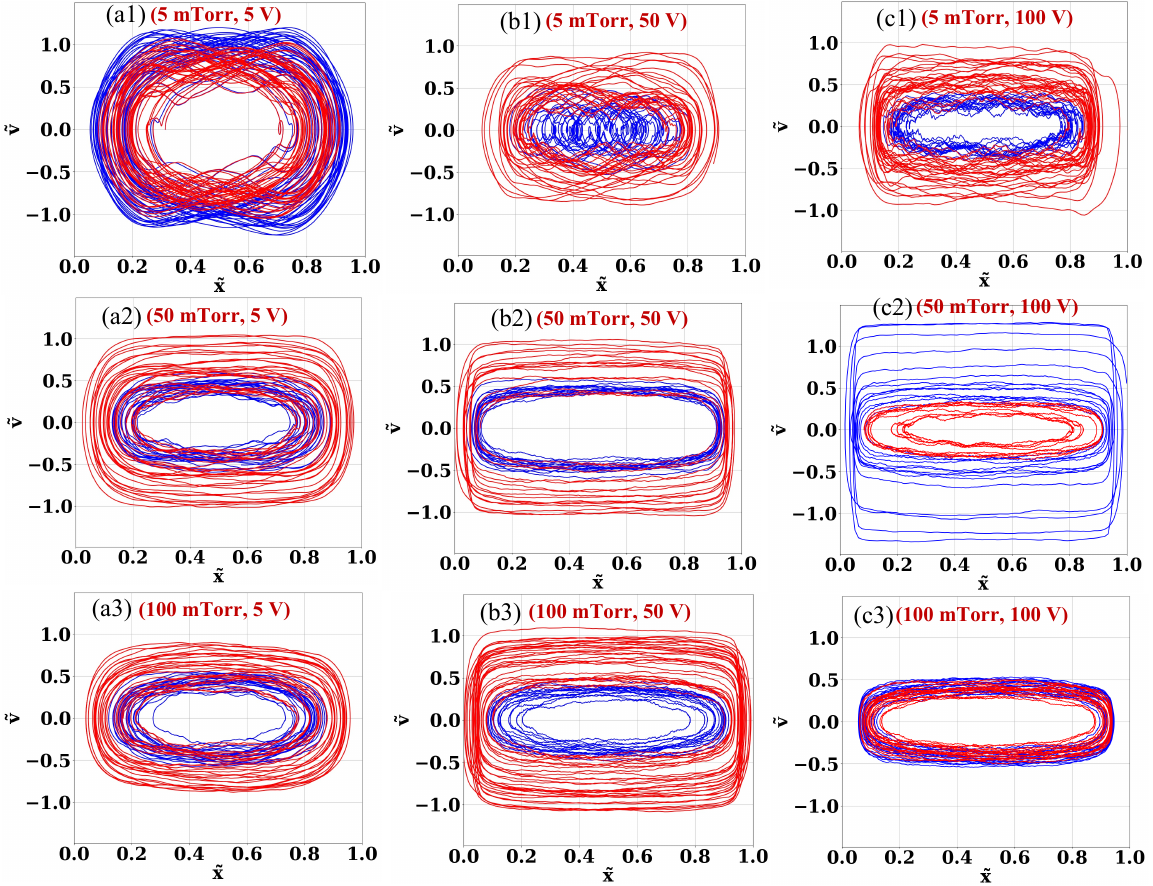}
  \caption{In this figure blue ($x_1$) and red ($x_2$) are phase space trajectories of 0 eV test electron for 100 rf cycle placed near the sheath, both are initially separated by $\lambda_{De} /120$ . (a1) to (a3) are at 5 V, (b1) to (b3) are at 50 V and (c1) to (c3) are at 100V for 5, 50 and 100 mTorr.}
 \label{ps}
\end{figure*}
\begin{figure*}
 \centering
 \includegraphics[width=\linewidth, height=11cm, keepaspectratio]{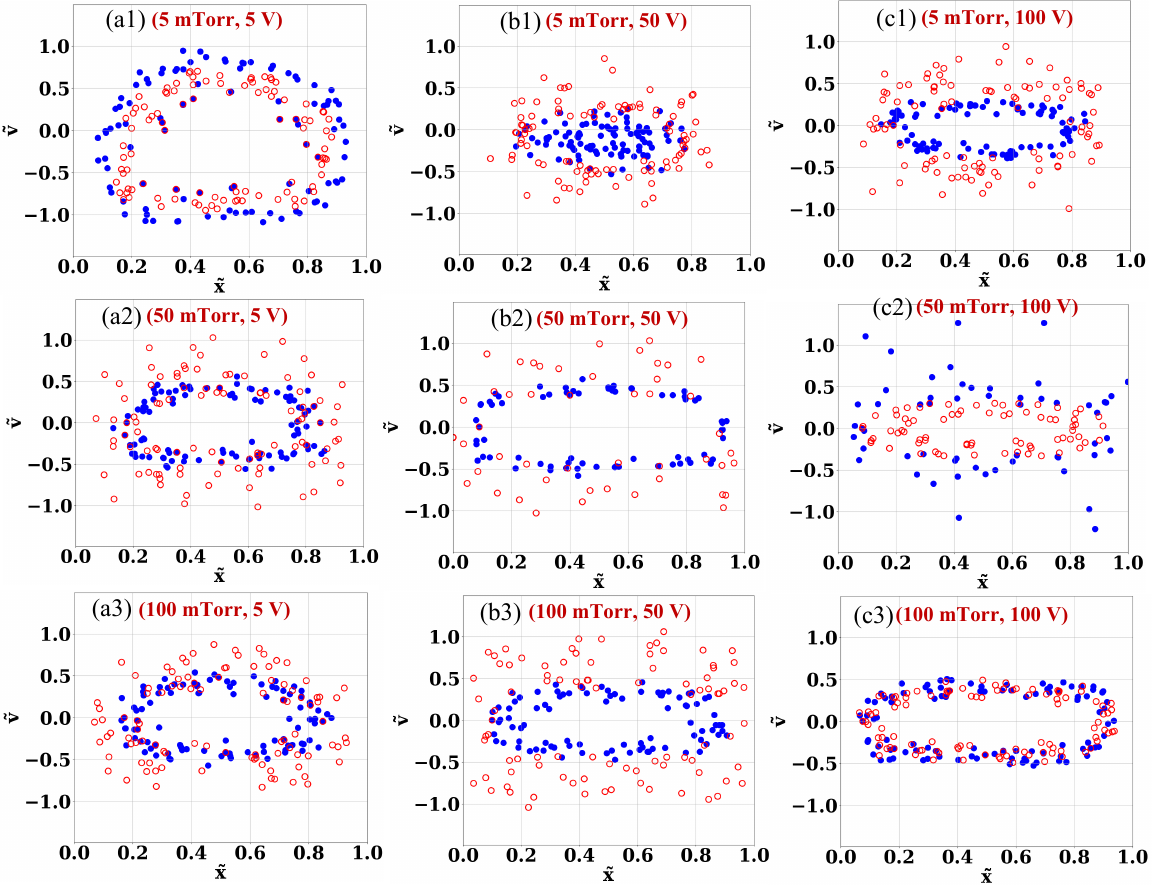}
  \caption{In this figure blue ($x_1$) and red ($x_2$) are stroboscopic view of phase space (Poincare section) taken at each rf cycle of 0 eV test electron for 100 rf cycle placed near the sheath, both are initially separated by $\lambda_{De} /120$ . (a1) to (a3) are at 5 V, (b1) to (b3) are at 50 V and (c1) to (c3) are at 100V for 5, 50 and 100 mTorr.}
 \label{pm}
\end{figure*}
\par A dynamical system is fundamentally characterized by its phase space, a multidimensional manifold that provides a comprehensive mathematical representation of the system’s instantaneous state. For the one-dimensional electron dynamics considered here, the phase space is defined by the electron’s position “x” and velocity “v” (or momentum p). Because the system is subject to a time-periodic radio-frequency (RF) electric field, the Hamiltonian becomes explicitly time-dependent, rendering the dynamics non-autonomous and increasing the effective dimensionality of the system.
\par Figure \ref{ps} presents the phase space plots of two test electrons — shown in blue (starting from position $x_1$) and red (starting from position $x_2$) — traced over 100 consecutive RF cycles following the attainment of steady-state conditions. Both electrons are initialized from rest (0 eV) and are placed in close proximity to the sheath edge, separated by an initial spatial displacement of  $\lambda_{De} /120$. The phase space coordinates are normalized: $\tilde{x}= x/L$ denotes the normalized spatial position within the discharge, where L is the electrode separation, and $\tilde{v}=v/(Lf_{rf})$ is the normalized electron velocity. The nine panels systematically cover the full parameter space: columns correspond to applied voltages of 5 V (panels a1–a3), 50 V (panels b1–b3), and 100 V (panels c1–c3), while rows correspond to neutral gas pressures of 5 mTorr (panels a1, b1, c1), 50 mTorr (panels a2, b2, c2), and 100 mTorr (panels a3, b3, c3).
\par At 5 mTorr (top row, panels a1 $\rightarrow$ b1 $\rightarrow$ c1), the phase space portrait at 5 V (panel a1) exhibits a nearly circular or elliptical loop structure, with the trajectories of both test electrons filling a broad annular region of phase space. The blue and red orbits are substantially interleaved with $0.4 \lesssim \lvert \tilde{v} \rvert \lesssim 1.2$, indicating that both electrons execute large-amplitude oscillations that traverse the full bulk plasma length. The prominent presence of filamentary structures inside the bulk plasma, seen in Fig.\ref{stp}(a1), result in distortions in the phase-space trajectories when $0.2 \lesssim \lvert \tilde{x} \rvert \lesssim 0.8$. Despite the presence of these nonlinear structures, the separation between initially neighboring orbits remains moderate over 100 RF cycles.
\par As the applied voltage is raised to 50 V (panel b1) and 100 V (panel c1)  at the same pressure of 5 mTorr, the red trajectories begin to spread more widely than the blue ones, with the outer loops of the red orbit extending to larger amplitudes while the blue orbits trace a comparatively tighter inner region. This divergence between the two orbits is a direct manifestation of the growing sensitivity to initial conditions: the increased nonlinearity of the electric field at higher voltage — evidenced by the more complex spatiotemporal transient structures discussed in the previous section — causes initially nearby trajectories to diverge at a faster rate.
\par At 50 mTorr (middle row, panels a2 → b2 → c2) and 100 mTorr (bottom row, panels a3 → b3 → c3), the overall trends are qualitatively similar but significantly influenced by increased electron-neutral collisions. Such collisions smoothen out the filamentary structures in the electric fields to which the test electrons respond. Therefore, at these pressures the filamentary structures in the electric field have either weakened (Fig.\ref{stp}(a2,a3)) or disappeared (Fig.\ref{stp}(b2,c2,b3,c3)). The disappearance of these electric field structures from bulk plasma is marked by formation of rectangular-shaped phase-space portrait as can be seen from panel b2 and c2 of Fig.~\ref{ps}. The rectangular phase-space portrait is indicative of electron velocity remaining unchanged inside the bulk plasma before it undergoes sharp reflection at the sheath. In contrast, the absence of rectangular phase-space indicates penetration of sheath fields. Interestingly, in panel c2 (100 V, 50 mTorr) the blue and red trajectories exhibit a striking role reversal relative to all other panels: the blue orbit now spans the outer, larger-amplitude region of phase space while the red orbit is confined to an inner, lower-amplitude loop. In contrast, in panel c3 (100 V, 100 mTorr) a near-perfect overlap between the blue and red orbits. Therefore collisional damping at 100 mTorr is sufficient to suppress the development of chaotic divergence even under conditions of strong RF driving voltage amplitude. 
\par To improve our understanding of stochastic dynamics of the electron inside the bulk plasma, we now extend the phase-space portraits picture through construction of Poincaré sections. 

\subsection{Poincaré sections}
A Poincaré section provides a simple way to visualize a high-dimensional flows by performing a dimensional reduction. For a periodically driven system, it is typically constructed by sampling the continuous phase-space trajectory at discrete temporal intervals $T=2\pi/\omega$, corresponding to the driving frequency $\omega$. For the problem under consideration, this involves recording the electron’s state (x,v) stroboscopically at a fixed phase of the RF cycle:
\begin{equation*}
 t_n =nT+\phi_0,	\,\,\,\,\,\,n=0,1,2,…   
\end{equation*}
This mapping transforms the continuous-time evolution into a discrete dynamical map, effectively filtering out the fast-scale oscillations to reveal the underlying topological structure of the phase space. This transition from a flow to a map allows for a rigorous distinction between:
\begin{itemize}
    \item Invariant Tori: Representing regular, quasi-periodic motion.
    \item Resonant Islands: Indicating phase-locking between the electron and the RF field.
    \item Stochastic Sea: Highlighting regions of chaotic motion where trajectories exhibit sensitive dependence on initial conditions.
\end{itemize}

In the specific domain of RF plasma discharges, this approach has been instrumental in understanding non-collisional heating. Goedde \textit{et~al.}\ \cite{Goedde1988} pioneered an analytical mapping approach, analogous to the Fermi-Ulam acceleration model, to investigate electron energy gain through interactions with oscillating RF sheath boundaries. Their findings revealed that the low-energy regions of phase space are often dominated by stochasticity, as repeated interactions with the sheath boundary lead to phase-randomization and subsequent "heating" through diffusion in the velocity space. By employing a surface-of-section method—sampling (x,v) at integer multiples of the RF period—they were able to quantify the boundaries of these stochastic regions. Lichtenberg \textit{et~al.}\ \cite{Lichtenberg1998} subsequently extended this framework to a broader parameter space in capacitively coupled plasmas (CCP), further establishing Poincaré mapping as a definitive diagnostic tool for characterizing the transition from regular to chaotic electron dynamics in low-pressure discharge environments. In the current studies, we extend this analysis to wider range of pressures and applied voltages by using trajectories of test particles obtained from PIC simulations. 

Figure \ref{pm} presents the Poincaré sections — equivalently, the stroboscopic phase space maps — of the same two test electrons whose continuous phase space trajectories were shown in Figure \ref{ps}. Each Poincaré section is obtained by retaining only those points from the corresponding continuous trajectory in Figure \ref{ps} that fall at the stroboscopic sampling instant — one point per RF period — discarding all intermediate trajectory information. As a consequence, regions of phase space that appeared densely filled with continuous loops in Figure \ref{ps} now reveal their underlying dynamical structure through the discrete point distributions of Figure \ref{pm}. The blue and red circles represent the normalized positions ($\tilde{x}$) and velocities ($\tilde{v}$) of these test electrons, sampled during every RF cycle over a fixed phase. The panel layout is identical to Figure 5: columns correspond to applied voltages of 5 V (panels a1–a3), 50 V (panels b1–b3), and 100 V (panels c1–c3), and rows correspond to neutral gas pressures of 5 mTorr (panels a1, b1, c1), 50 mTorr (panels a2, b2, c2), and 100 mTorr (panels a3, b3, c3). 

\par Consistent with phase-space portrait, depicted in Fig.~\ref{ps}, we observe that the electron motion becomes increasingly stochastic with increasing applied voltage at 5 mTorr (top row, panels a1 → b1 → c1). At 5 V (panel a1), both the test electrons undergo quasi-periodic motion as can be seen by elliptical orbits in the Poincaré section. However, with increasing voltage (panel b1, c1) scattering of red dots indicate that the corresponding electron sees random phase of oscillations each time it returns to the sheath. Similar trend continues for 50 mTorr case (middle row, panels a2 → b2 → c2), where increasing applied voltage makes the electron motion increasingly stochastic. Interesting, this trend is reverse for 100 mTorr case (bottom row, panels a3 → b3 → c3), where we observe both electrons undergoing quasi-periodic motion at 100 V (panel c3).  
\par Thus, from the analysis of phase-space portrait and Poincaré sections of test electrons we conclude that  with increasing applied voltage the stochasticity of electrons, located near the sheath region, increases at low pressure (5 mTorr) conditions and decreases at high pressure (100 mTorr) conditions. The increase in stochasticity with increasing applied voltage at 5 mTorr pressure is mainly due to enhanced non-linearity in the electric field structures penetrating inside the bulk plasma. In contrast, breaking of filamentary structures due to enhance electron-neutral collisions smoothens the electric field inside the bulk plasma at high voltage, high pressure conditions leading to reduction in stochasticity. 
\par Although phase-space portrait and Poincaré sections provide us qualitative trends about stochastic motion of electrons inside the bulk plasma, they are of limited use in making quantitative predictions. In order to quantify the degree of stochasticity in electron motion, we now analyze the trajectories in terms of the Lyapunov exponents associated with them.  

\subsection{Lyapunov Exponent}  
\begin{figure*}
 \centering
 \includegraphics[width=\linewidth, height=10 cm, keepaspectratio]{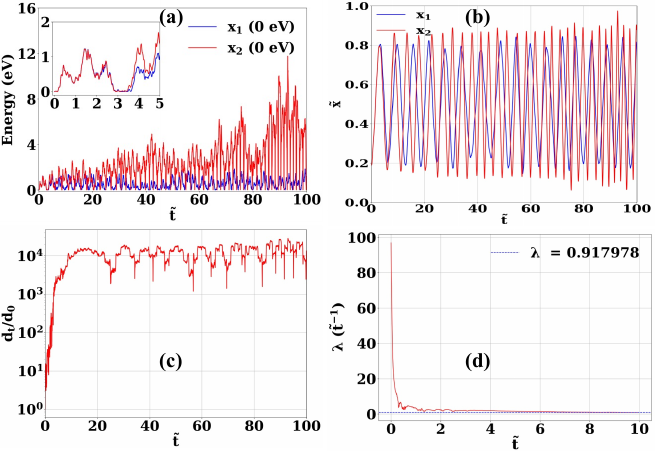}
  \caption{This Figure illustrates the chaotic dynamics of two electrons initialized with 0 eV kinetic energy at nearly identical positions in the bulk plasma, separated by $\lambda_{De}/120$ at 5 mTorr and 100 V. Panel (a) shows the temporal evolution of the kinetic energy of both electrons over 100 RF cycles; the electron starting form position $x_1$ (blue) remains at low energy throughout, while electron at initial position $x_2$ (red) undergoes stochastic heating and reaches energies exceeding 4 eV. Panel (b) displays the normalized position trajectories $\tilde{x}$ of both electrons as a function of dimensionless time $\tilde{t}$, revealing divergent spatial behavior — the electron starting from $x_2$ repeatedly approaches and interacts with the sheath edge, while other electron remains confined to the bulk. Panel (c) presents the normalized phase-space separation $d_t/d_0$ between the two electron trajectories on a logarithmic scale over 100 RF cycles, demonstrating exponential growth followed by saturation near $\approx10^4$. Panel (d) shows the time-evolving Lyapunov exponent $\lambda$ computed over the initial 10 RF cycles within the linear-growth regime, converging to a positive value of $\lambda = 0.917978 \hspace{.1 cm} \tilde{t}^{-1}$, confirming the chaotic nature of electron motion.}
 \label{lp}
\end{figure*} 
The Lyapunov exponent is a fundamental quantity in nonlinear dynamics that provides a quantitative measure of the sensitivity of a dynamical system to its initial conditions and serves as one of the most widely used indicators of chaotic behavior \cite{bak, strogatz}. In nonlinear systems, two trajectories that begin from nearly identical initial states may evolve in dramatically different ways as time progresses. Depending on the nature of the underlying dynamics, the trajectories may remain close to one another, converge toward a stable orbit, or diverge rapidly in phase space. 
\par Physically, this behavior implies that the long-term evolution of the system becomes extremely sensitive to microscopic variations in the initial state. As a result, two particles subjected to nearly identical initial conditions can eventually follow entirely different trajectories. This sensitive dependence on initial conditions is regarded as one of the defining signatures of deterministic chaos. Consequently, the Lyapunov exponent provides an effective tool for distinguishing ordered dynamics from stochastic or chaotic motion in nonlinear plasma systems.
\par In the context of low-pressure RF capacitively coupled plasmas, the electron motion is governed by strongly nonlinear interactions with oscillating sheath electric fields, transient electric fields inside the bulk plasma and collisionless wave–particle interactions. These interactions can generate stochastic trajectories in phase space, particularly when the electron bounce dynamics become coupled to the oscillatory RF sheath motion. Therefore, analyzing the sensitivity of nearby electron trajectories offers important insight into the stochastic heating and energy transport of electrons in CCP discharges.
\par To quantify this behavior, we examine the temporal evolution of two initially nearby test electron trajectories in phase space. If the initial phase-space separation between the two trajectories is $d_0$, then their separation after time $t$ can be expressed as
\begin{equation}
		d_t \approx d_0 e^{\lambda t}
	\end{equation}
where $d_t$ is the instantaneous distance between the trajectories and $\lambda$ is the Lyapunov exponent. A positive value of $\lambda$ ($\lambda>0$) indicates exponential divergence of trajectories and therefore chaotic dynamics, whereas negative values ($\lambda$<0) correspond to stable motion where trajectories converge toward an attractor. A Lyapunov exponent close to zero ($\lambda=0$) represents marginal stability, commonly associated with periodic or quasiperiodic motion. Thus, the Lyapunov exponent provides a direct quantitative measure of the degree of stochasticity and nonlinear sensitivity present in the electron dynamics. In the present study, it is employed to characterize the chaotic nature of electron oscillations inside the bulk plasma and to investigate how this behavior evolves with variations in discharge voltage and neutral gas pressure.
Taking the logarithm of equation (2) gives\cite{ueshima1997}
\begin{equation}
		\lambda \approx \frac{1}{t} \ln\left( \frac{d_t}{d_0} \right)
\end{equation}
For numerical calculations over discrete time intervals, the Lyapunov exponent can be expressed as 
    \begin{equation}
    \lambda \approx \frac{1}{t} \sum_{i=1}^{N} \ln\left( \frac{d_i}{d_{i-1}} \right)
    \label{lp_num}
    \end{equation}
where $d_i$ represents the phase-space distance between the two trajectories at the sampling time $i \Delta t$. 
In the present work, the phase-space separation $d_i$ between two test electrons is calculated as:
\begin{equation}
d_i = \sqrt{(x_1(i) - x_2(i))^2 + (v_1(i) - v_2(i))^2}
\label{ps_sep}
\end{equation}
where $x_1(i)$ and $x_2(i)$ are the instantaneous positions, and $v_1(i)$ and $v_2(i)$ are the corresponding velocities at the $i^{th}$ time step.
Since both test electrons start from rest, the initial phase-space separation $d_0$ equal to the initial spatial separation i.e. $ d_0 = \lambda_{De}/120$. 

\begin{figure*}
 \centering
\includegraphics[width=\textwidth]{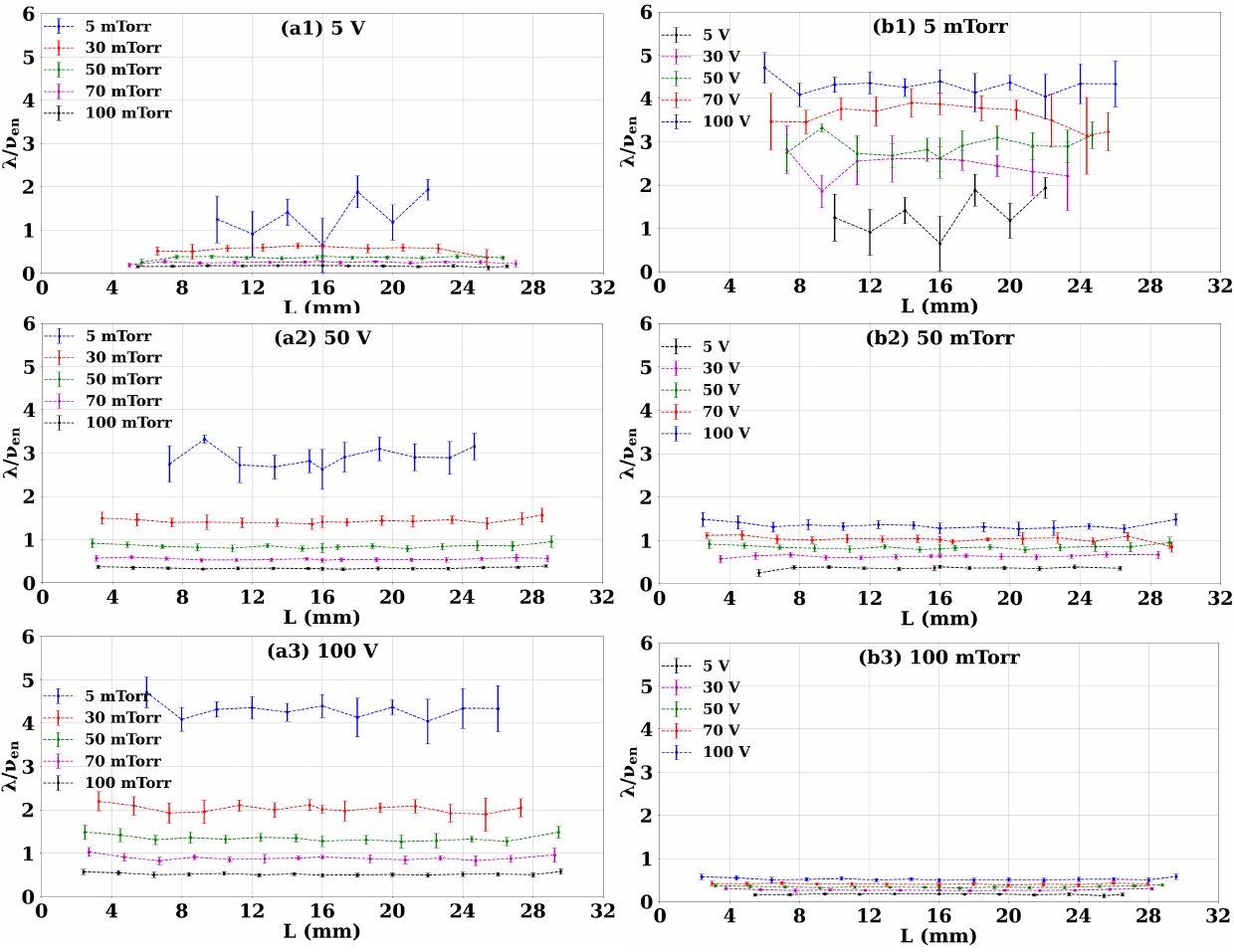}
  \caption{This Figure presents the spatial variation of the normalized Lyapunov exponent $\lambda/\nu_{en}$ across the bulk plasma, from the left sheath edge to the right sheath edge, where $\lambda$ is the maximal Lyapunov exponent characterizing the rate of chaotic trajectory divergence and $\nu_{en}$ is the elastic electron–neutral collision frequency serving as the collisional reference rate. Each data point represents the mean value (solid marker) of $\lambda/\nu_{en}$ computed over the 15 sub-interval realizations within each spatial bin $\Delta x$ of width 2 mm, and the error bars denote the corresponding standard deviation, reflecting the spatial variability of chaotic strength within each bin. The left column, panels (a1), (a2), and (a3), shows results at fixed applied voltages of 5 V, 50 V, and 100 V respectively, with pressure as the varying parameter: 5 mTorr (blue), 30 mTorr (red), 50 mTorr (green), 70 mTorr (magenta), and 100 mTorr (black). The right column, panels (b1), (b2), and (b3), shows results at fixed pressures of 5 mTorr, 50 mTorr, and 100 mTorr respectively, with applied voltage as the varying parameter: 5 V (black), 30 V (magenta), 50 V (green), 70 V (red), and 100 V (blue).}
 \label{mapl}
\end{figure*} 
The procedure to compute Lyapunov exponential is illustrated with the help of Fig.~\ref{lp} for case of 5mTorr and 100 V. Here, the kinetic energy evolution of two electrons, both initialized with 0 eV, is shown in the Fig.~\ref{lp}(a). The two electrons are initially placed at locations $x_1 = 6$ mm and $x_2 = 6 + d_0$ mm. Despite this negligibly small initial separation, the two trajectories diverge dramatically over time, demonstrating the case of deterministic chaos. By the end of 100 RF cycles, the electron starting from $x_2$ has undergone stochastic motion, accumulating maximum kinetic energy up to 12 eV, while the electron starting from $x_1$ gains comparatively very little energy throughout the simulation. This asymmetric energy gain, arising from an arbitrarily small initial perturbation, is a direct manifestation of sensitive dependence on initial conditions — a defining characteristic of chaotic dynamical systems. Further, from Fig.~\ref{lp}(b) we see that the electron starting from $x_2$ (shown in red color) makes repeated excursions deep inside the sheath region. Therefore, due to repeated interaction with the oscillating sheath electric field, this electron gains significantly more energy compared to the one starting from position $x_1$ (shown in blue color). 
\par The Fig.~\ref{lp}(c) quantifies the phase-space divergence, calculated with Eq.~\ref{ps_sep}, between the two trajectories by plotting the time evolution of the ratio $d_t/d_0$ on a logarithmic scale over 100 RF cycles. The separation grows exponentially during approximately the first 10 RF cycles, reaching a value of $\approx10^4$ times the initial separation, before saturating due to the finite size of the available phase space. This exponential growth phase is consistent with chaotic dynamics, where nearby trajectories diverge at a rate governed by the Lyapunov exponent. The subsequent saturation is a natural consequence of the bounded nature of the system: once the two trajectories have fully decorrelated and are sampling different regions of the chaotic attractor, no further relative divergence is possible.
\par The instantaneous estimate of $\lambda$, defined by Eq.~\ref{lp_num}, is shown in Fig.~\ref{lp}(d). This shows that the Lyapunov exponent converges smoothly to a positive asymptotic value of $\lambda = 0.917978 \hspace{.1 cm} \tilde{t}^{-1}$ in normalized units. The positivity of the maximal Lyapunov exponent is the formal confirmation of chaos. Converting to physical units using the RF frequency $f_{rf}$, the Lyapunov exponent becomes $\lambda = 0.917978 \times f_{rf} \simeq 5.5 \times 10^{7} \hspace{.1 cm} \mathrm{s}^{-1}$, indicating that phase-space trajectories decorrelate on a timescale of order $1/\lambda \approx$ 18 ns, comparable to a period of applied RF voltage. This validates the use of stochastic heating models to describe electron energy gain in capacitively coupled plasma discharges.

\par We now extend this analysis to systematically characterize the spatial dependence and robustness of such chaotic behavior under various discharge operating conditions such as background neutral gas pressure, applied RF voltage, RF frequency etc. To characterize spatial dependence of the Lyapunov exponent under given operating conditions, we initialize test-particle electrons at different positions throughout the bulk plasma. Around a given spatial location, inside the bulk plasma, we introduce 15 pairs of test electrons having separation $\lambda_{De} /120$. This yields 15 independent estimates of the Lyapunov exponent at each spatial location, capturing the local variability of chaotic behavior due to the fine structure of the electric field and the sensitivity of trajectories to their precise starting point. The mean and standard deviation of the Lyapunov exponent are then computed across the 15 realizations at each spatial location, providing both a representative estimate of the local chaos strength and a quantitative measure of its spatial variability. This procedure is then repeated systematically across the entire bulk plasma, tracing the full spatial extent from one sheath edge to the other, thereby constructing a spatially resolved map of chaotic electron dynamics under the given discharge conditions.  
\par To place the degree of chaos in a physically meaningful context, each computed Lyapunov exponent $\lambda$ is normalized by the elastic electron–neutral collision frequency $\nu_{en}$, which represents the rate at which electrons lose coherence due to collisional scattering with the background neutral gas. The dimensionless ratio $\lambda/\nu_{en}$ thus provides a direct comparison between the rate of chaotic (collisionless, field-driven) trajectory divergence and the rate of collisional randomization. When $\lambda/\nu_{en} > 1$, the chaotic dynamics dominate over collisional effects, indicating that dominance of stochastic heating over collisional heating. Conversely, when $\lambda/\nu_{en} \approx 1$ , collisional and chaotic processes are comparable in importance. Finally, for dominance of collisional heating over stochastic heating, we expect $\lambda/\nu_{en} < 1$. 
\par Figure \ref{mapl} presents a comprehensive spatially resolved characterization of the normalized Lyapunov exponent $\lambda/\nu_{en}$ across the full extent of the bulk plasma, spanning from the left sheath edge to the right sheath edge with a spatial resolution of 2 mm. At each location, the mean (solid marker) and standard deviation (error bar) of $\lambda/\nu_{en}$ are computed from the ensemble of 15 test-particle pairs initialized symmetrically around that location, as described previously. The ratio $\lambda/\nu_{en}$ is the central diagnostic quantity of this analysis, decreasing transition from stochastic to collisional heating in CCP discharges. One of the striking feature of this analysis is the uniformity of normalized Lyapunov exponent across the bulk plasma.           
\par The panels (a1) through (a3) of Fig.~\ref{mapl} display $\lambda/\nu_{en}$ as a function of position at fixed applied voltages of 5 V, 50 V, and 100 V respectively, with neutral gas pressure as the parametric variable. Similarly, panels (b1) through (b3) examine the complementary perspective, plotting $\lambda/\nu_{en}$ at fixed pressures of 5 mTorr, 50 mTorr, and 100 mTorr respectively, with applied voltage as the parametric variable. From this figure, we see that at an applied voltage of 5 V, the electron heating largely remains collisional as normalized $\lambda$ remains in the range $0<\lambda/\nu_{en}<2$. But with increase in applied voltage, the stochastic heating starts dominating over collisional heating. For example, at applied voltages of 50 V and 100 V, we see that $\lambda/\nu_{en}$ remains above unity up to operating pressures of 30 mTorr and 50 mTorr, respectively. Therefore, with increasing applied voltage stochastic heating becomes effective even at higher pressure as due to efficient randomization of electrons due to stronger sheath fields. At a fixed gas pressure, in general we see that $\lambda/\nu_{en}$ also increases with the applied RF voltage. We observe that stochastic heating is dominant only at low pressure conditions of 5 mTorr. The large error bars visible at low pressure and low voltage reflect the heightened spatial inhomogeneity of the chaotic dynamics under these conditions. Apart from 5 mTorr and 100 V case, $\lambda/\nu_{en}$ is observed to be broadly uniform across the bulk in most cases, with no strong systematic gradient from the left sheath edge to the right sheath edge. This spatial homogeneity suggests that the chaotic heating mechanism, while originating from sheath-edge interactions, imprints a relatively uniform degree of stochasticity throughout the bulk plasma on the timescale of many RF cycles, consistent with the global nature of stochastic heating in symmetric capacitively coupled plasma discharges. Therefore, it is possible to characterize the stochastic behavior of electrons under a given operating condition with a mean of normalized Lyapunov exponents, computed across the bulk plasma. 
\begin{figure}[!htbp]
 \centering
 \includegraphics[width=92.5 mm, height=77 mm ]{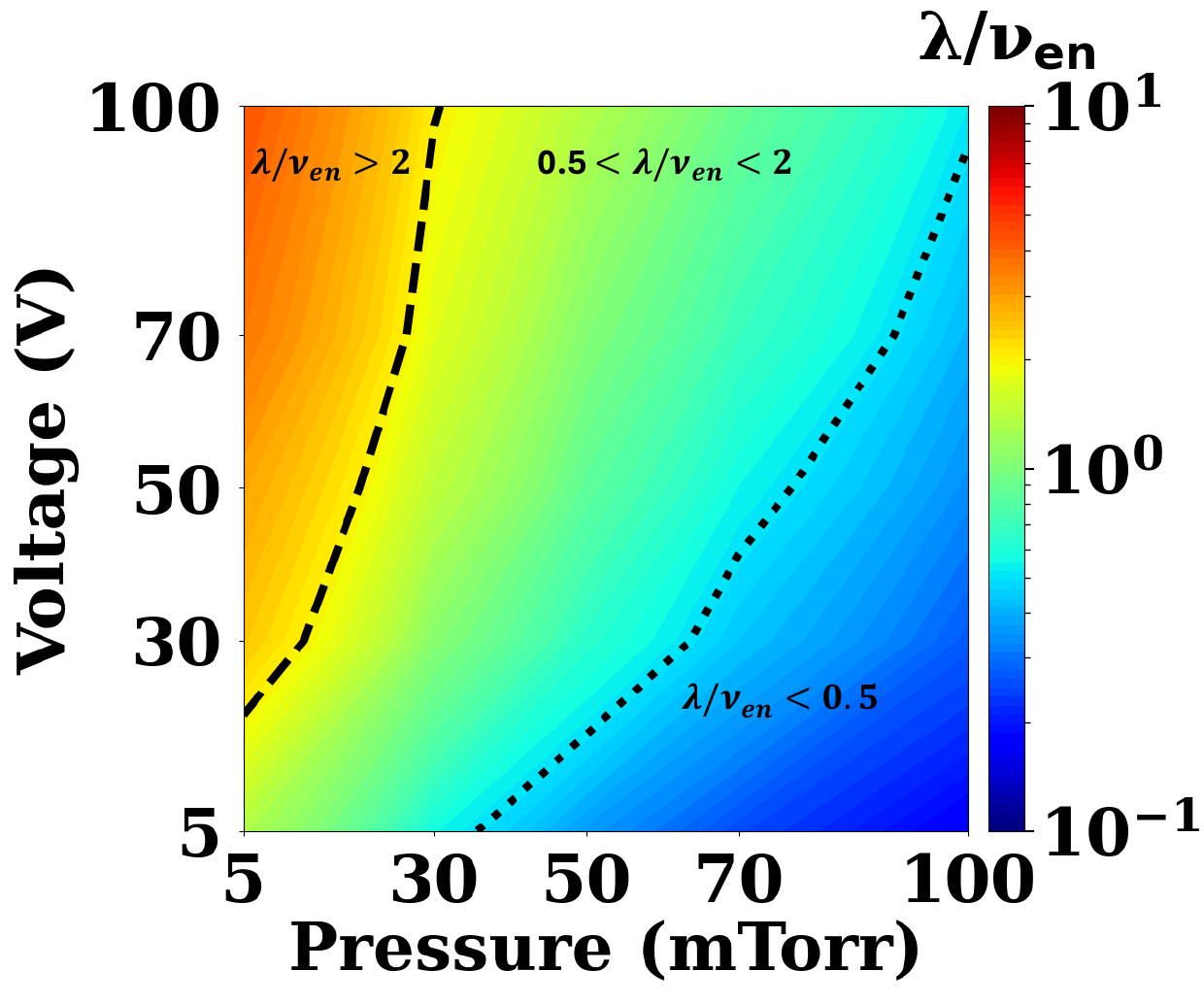}
  \caption{This Figure presents a two-dimensional color map of the spatially averaged normalized Lyapunov exponent $\lambda/\nu_{en}$ as a function of both applied RF voltage (5-100 V, vertical axis) and neutral gas pressure (5–100 mTorr, horizontal axis). This phase-diagram demonstrates the transition from stochastic to collisional regime of CCP having discharge length 32 mm and operating frequency 60 MHz. Here, strongly stochastic regime (red color) is characterized by $\lambda/\nu_{en} > 2$. On the other hand, collisional regime (blue color) is defined by $\lambda/\nu_{en} < 0.5$. Finally, the transition from strongly stochastic to collision regime happens at intermediate range (green color) $0.5 < \lambda/\nu_{en} < 2$.}
 \label{vpt}
\end{figure} 
\section{Discussion} 
\begin{figure*}[!htbp]
 \centering
 \includegraphics[width=\linewidth]{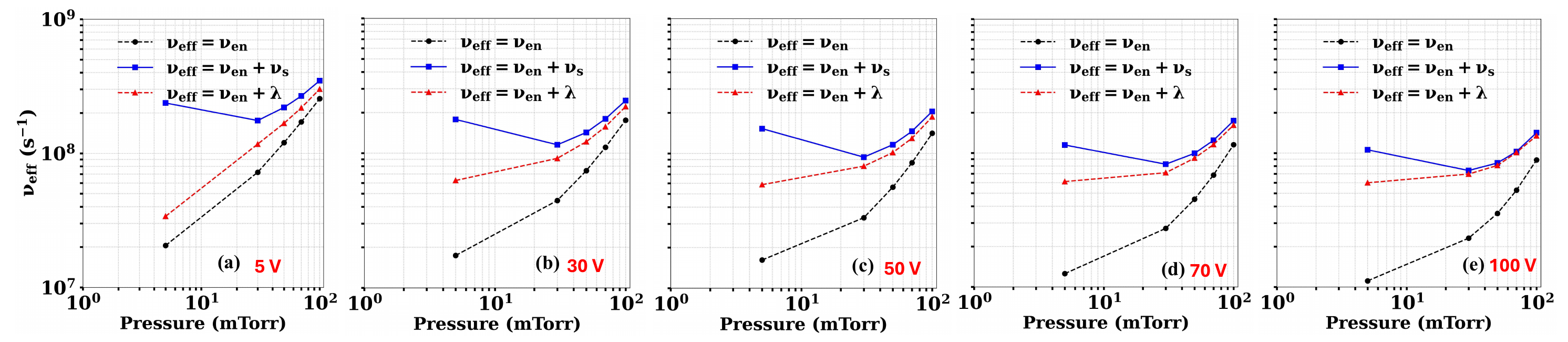}
  \caption{This Figure presents a comparison of three formulations of the effective electron collision frequency $\nu_{eff}$ as a function of neutral gas pressure (5-100 mTorr, on a logarithmic scale) at a fixed applied voltage of (a) 5 V, (b) 30 V, (c) 50 V, (d) 70 V and (e)100 V, along with the corresponding pressure dependence of the normalized Lyapunov exponent. Panel (a) shows $\nu_{eff}$ on a log-log scale for three cases: (i) $\nu_{eff} = \nu_{en}$  (black dashed, solid circles), representing the bare elastic electron–neutral collision frequency alone; (ii) $\nu_{eff} = \nu_{en} + \nu_s$ (blue solid, solid squares), where $\nu_s = 2\bar{\mathrm{v}}/{L_0}$  is the Godyak stochastic collision frequency accounting for collisionless energy exchange between electrons and the oscillating sheaths; and (iii) $\nu_{eff} = \nu_{en} + \lambda$ (red dashed, solid triangles), where the Lyapunov exponent $\lambda$ replaces $\nu_s$ as the measure of collisionless heating rate, providing a nonlinear dynamical characterization of stochastic electron–sheath interactions.} 
 \label{tran}
\end{figure*} 
\par The variation of normalized Lyapunov exponent $\lambda/\nu_{en}$ over the entire operating parameter space reveals the nature of CCP discharge for a given discharge length and operating frequency. For a specific case considered in the present study, i.e. discharge length $L =$ 32 mm and operating frequency of 60 MHz, this variation is summarized in the form of a phase diagram shown in the Fig.~\ref{vpt}. Here, we have categorized the operating conditions into three distinct regimes viz. strongly stochastic regime (red color), mildly stochastic regime (green color) and collisional regime (blue color). The strongly stochastic regime is defined with the condition, $\lambda/\nu_{en} > 2$. This regime (red color) is accessible in the pressure range 5 mTorr to 10 mTorr at applied voltage of 30 V, whereas the range expands to 30 mTorr at 100 V applied voltage. In contrast, the collisional regime (blue color) is characterized by the condition $\lambda/\nu_{en} < 0.5$. In this case, the heating is predominantly collisional at the applied voltages of 5V when pressure exceed 30 mTorr. Similarly, at 100 V applied voltage, collisional heating happens at the pressure of 100 mTorr. Finally, in the intermediate range $0.5 < \lambda/\nu_{en} < 2$, the transition between stochastic heating to collisional heating occurs. 

\par These findings are consistent with and extend a body of prior theoretical and experimental work on the pressure-dependent transition between stochastic and ohmic heating regimes in capacitively coupled plasmas. Popov \textit{et al.}\ \cite{popov1985power} experimentally studied a CCP discharge in mercury vapor over a pressure range of $0.1$--$10$  mTorr and at applied rf voltage below 200 V, and demonstrated that, below approximately $10$ mTorr, the RF power dissipated through electron–sheath interactions exceeded over dissipation through electron–neutral collisions in the bulk, directly establishing the low-pressure dominance of collisionless heating. Godyak \cite{godyak1986soviet} introduced a theoretical framework for this transition by defining an effective collision frequency in the form
\begin{equation}
  \nu_{eff} = \nu_{en} + \nu_s; \,\,\,\,\,\,\, \nu_s = 2\bar{\mathrm{v}}/{L_0}, 
  \label{nu_eff}
\end{equation}
where $\nu_s$ is the stochastic collision frequency, which characterizes the rate of electron–sheath energy exchange through mean electron thermal speed $\bar{\mathrm{v}} = \sqrt{8eT_e/\pi m}$ and plasma thickness $L_0$. This expression makes explicit that stochastic heating acts as an effective collisional process, contributing to electron energy dissipation even in the complete absence of real collisions, and that its relative importance grows with increasing electron temperature and decreasing plasma length, both conditions favoring low pressure and high voltage. Godyak \cite{godyak1986soviet} further demonstrated for argon discharges that the ratio of stochastic to ohmic power $P_{st}/P_\nu$ decreases with increasing pressure, in direct agreement with the trend observed here in $\lambda/\nu_{en}$. These experimental findings were subsequently validated by Lafleur \textit{et al.}\cite{lafleur2013anomalous,lafleur2014electron} using fully kinetic PIC simulations, which confirmed the pressure-dependent transition and provided detailed spatial resolution of the power deposition profiles associated with each heating mechanism. The present Lyapunov-based analysis provides a complementary and fundamental characterization of this same transition, in a rigorous language of nonlinear dynamics. 
\par Now we compare our results with Godyak's theoretical estimates for stochastic collisional frequency $\nu_s$, defined in Eq.~\ref{nu_eff}. In our case, the stochastic collision frequency is taken to be equal to the numerically computed Lyapunov exponent. Therefore, the effective collision frequency takes the form $\nu_{eff} = \nu_{en} + \lambda$. Fig.~\ref{tran} provides a quantitative comparison between three physically motivated definitions of the effective electron collision frequency $\nu_{eff}$, evaluated as a function of neutral gas pressure ranging from 5 to 100 mTorr at various fixed applied RF voltages. Here, black curves represent purely collisional estimates i.e. $\nu_{eff} = \nu_{en}$. The blue and red curves are plotted with stochastic collision frequency is considered to equal to $2\bar{\mathrm{v}}/{L_0}$ and Lyapunov exponents ($\lambda$), respectively. The thermal velocity $\bar{\mathrm{v}}$ is calculated using the electron temperature $T_e$, extracted from the electron energy distribution function (EEDF) evaluated self-consistently at the center of the discharge from the PIC simulation at steady state. Also, $L_0$ is taken from bulk plasma length given in the table ~\ref{tab}. From Fig.~\ref{tran}, we see that at higher pressures (above 50 mTorr), stochastic collision frequency given by Godyak's model matches well with estimates obtained from the Lyapunov exponent. In particular, this agreement is better at higher applied voltage (see Fig.~\ref{tran}(c),(d),(e)). As we move to the lower pressure cases, the two estimates start deviating from each other. In general, Godyak's estimate of stochastic frequency is found to be higher than the corresponding Lyapunov exponent. The difference becomes significant at extremely low pressure of 5 mTorr, with discrepancy gradually increasing with decreasing applied voltage. This is primarily due to reduction in sheath electric field with decreasing applied voltage. Due to this reduction, the electron scattering with sheath interaction is significantly reduced in lower voltage, low pressure (LVLP) discharges. In other words, in LVLP conditions electrons experience a slowly varying electric field instead of strongly localized sheath electric field, typically responsible for their stochastic motion. The phenomenological model does not capture this dynamics, causing overestimation of stochastic frequency under LVLP conditions. From this discussion, we conclude that the Lyapunov exponent offers a rigorous and independent measure of the collisionless heating rate that is quantitatively consistent with the classical stochastic frequency framework of Godyak, \cite{godyak1986soviet,Godyak1990} while providing a fundamental dynamical basis rooted in nonlinear chaos theory rather than in phenomenological kinetic arguments.  
 
 \section{Conclusion}  
In this work, we present a comprehensive analysis of stochastic motion of an electrons in CCP discharges using three non-linear dynamical tools, namely phase-space portrait, Poincaré section and the Lyapunov exponent. While phase-space portrait and Poincaré section analysis established the stochasticity of electrons under wide range of operating conditions, the analysis with Lyapunov exponent provided quantitative measure. The origin of stochastic stems from repeated  interactions of electrons with the sheath electric field. Additionally, transient filamentary electric fields penetrating inside the bulk plasma, particularly at lower pressures, contribute to the randomization of electron motion. Typically, we observe that stochasticity increases with increasing applied voltage due to enhancement in the sheath electric field. In contrast, the electric field inside the bulk becomes smoothened leading to descrease in stochasticity with increasing pressure. 
\par Phase-diagram, constructed with normalized Lyapunov exponent $\lambda/\nu_{en}$, categorizes the operation of CCPs in three distinct regimes, viz. strongly stochastic ($\lambda/\nu_{en} > 2$), mildly stochastic ($0.5 < \lambda/\nu_{en} < 2$) and collisional ($\lambda/\nu_{en} < 0.5$) regime. This diagram provides an effect tool to distinguish collisional heating from stochastic heating under given operating conditions such as neutral gas pressure and applied RF voltage. Additionally, the paper conclusively demonstrates use of Lyapunov exponent as an effective stochastic collision frequency and compares it with earlier theoretical estimates, which were based on phenomenological model. The present analysis thus establishes that the Lyapunov exponent provides not only a qualitative indicator of chaos, but a quantitatively meaningful and physically interpretable measure of the collisionless heating rate in capacitively coupled RF plasma discharges, directly comparable to experimentally and computationally established benchmarks from the literature.  

\begin{acknowledgments}
The simulation work presented here is
performed on ANTYA cluster at the Institute for Plasma Research (IPR), Gandhinagar, India.
\end{acknowledgments}
\section*{AUTHOR DECLARATIONS}
\section*{Conflict of Interest}
The authors have no conflicts to disclose.
\section*{Data Availability}
The data that support the findings of this study are available
from the corresponding author upon reasonable request.
\nocite{*}
\section*{references}
\bibliography{aipsamp}
\end{document}